\documentclass[twocolumn,aps,showpacs,superscriptaddress]{revtex4}
\usepackage{graphicx}

\newcommand{ \pp }{$p$ + $p$ }
\newcommand{\srt}{$\sqrt{s}$ }

\newcommand{\met}{$\langle E_T \rangle$ }
\newcommand{\pt}{$p_T$ }

\newcommand{\Dstar}{$D^{*}$ }
\newcommand{\Dstarpm}{$D^{*\pm}$ }
\newcommand{\Nratio}{$N(D^{*+}+D^{*-})/N(\mathrm{jet})$ }

\begin{document}
\def\Journal#1#2#3#4{{#1} {\bf #2}, #3 (#4)}

\def\NCA{Nuovo Cimento}
\def\NIM{Nucl. Instr. Meth.}
\def\NIMA{{Nucl. Instr. Meth.} A}
\def\NPB{{Nucl. Phys.} B}
\def\NPA{{Nucl. Phys.} A}
\def\PLB{{Phys. Lett.}  B}
\def\PRL{{Phys. Rev. Lett.}}
\def\PRC{{Phys. Rev.} C}
\def\PRD{{Phys. Rev.} D}
\def\ZPC{{Z. Phys.} C}
\def\JPG{{J. Phys.} G}
\def\EPJ{{Eur. Phys. J.} C}
\def\RPP{{Rep. Prog. Phys.}}

\preprint{}
\title{Measurement of \Dstar Mesons in Jets from \pp Collisions at \srt = 200 GeV}

\affiliation{Argonne National Laboratory, Argonne, Illinois 60439, USA}
\affiliation{University of Birmingham, Birmingham, United Kingdom}
\affiliation{Brookhaven National Laboratory, Upton, New York 11973, USA}
\affiliation{University of California, Berkeley, California 94720, USA}
\affiliation{University of California, Davis, California 95616, USA}
\affiliation{University of California, Los Angeles, California 90095, USA}
\affiliation{Universidade Estadual de Campinas, Sao Paulo, Brazil}
\affiliation{University of Illinois at Chicago, Chicago, Illinois 60607, USA}
\affiliation{Creighton University, Omaha, Nebraska 68178, USA}
\affiliation{Nuclear Physics Institute AS CR, 250 68 \v{R}e\v{z}/Prague, Czech Republic}
\affiliation{Laboratory for High Energy (JINR), Dubna, Russia}
\affiliation{Particle Physics Laboratory (JINR), Dubna, Russia}
\affiliation{Institute of Physics, Bhubaneswar 751005, India}
\affiliation{Indian Institute of Technology, Mumbai, India}
\affiliation{Indiana University, Bloomington, Indiana 47408, USA}
\affiliation{Institut de Recherches Subatomiques, Strasbourg, France}
\affiliation{University of Jammu, Jammu 180001, India}
\affiliation{Kent State University, Kent, Ohio 44242, USA}
\affiliation{University of Kentucky, Lexington, Kentucky, 40506-0055, USA}
\affiliation{Institute of Modern Physics, Lanzhou, China}
\affiliation{Lawrence Berkeley National Laboratory, Berkeley, California 94720, USA}
\affiliation{Massachusetts Institute of Technology, Cambridge, MA 02139-4307, USA}
\affiliation{Max-Planck-Institut f\"ur Physik, Munich, Germany}
\affiliation{Michigan State University, East Lansing, Michigan 48824, USA}
\affiliation{Moscow Engineering Physics Institute, Moscow Russia}
\affiliation{City College of New York, New York City, New York 10031, USA}
\affiliation{NIKHEF and Utrecht University, Amsterdam, The Netherlands}
\affiliation{Ohio State University, Columbus, Ohio 43210, USA}
\affiliation{Old Dominion University, Norfolk, VA, 23529, USA}
\affiliation{Panjab University, Chandigarh 160014, India}
\affiliation{Pennsylvania State University, University Park, Pennsylvania 16802, USA}
\affiliation{Institute of High Energy Physics, Protvino, Russia}
\affiliation{Purdue University, West Lafayette, Indiana 47907, USA}
\affiliation{Pusan National University, Pusan, Republic of Korea}
\affiliation{University of Rajasthan, Jaipur 302004, India}
\affiliation{Rice University, Houston, Texas 77251, USA}
\affiliation{Universidade de Sao Paulo, Sao Paulo, Brazil}
\affiliation{University of Science \& Technology of China, Hefei 230026, China}
\affiliation{Shandong University, Jinan, Shandong 250100, China}
\affiliation{Shanghai Institute of Applied Physics, Shanghai 201800, China}
\affiliation{SUBATECH, Nantes, France}
\affiliation{Texas A\&M University, College Station, Texas 77843, USA}
\affiliation{University of Texas, Austin, Texas 78712, USA}
\affiliation{Tsinghua University, Beijing 100084, China}
\affiliation{United States Naval Academy, Annapolis, MD 21402, USA}
\affiliation{Valparaiso University, Valparaiso, Indiana 46383, USA}
\affiliation{Variable Energy Cyclotron Centre, Kolkata 700064, India}
\affiliation{Warsaw University of Technology, Warsaw, Poland}
\affiliation{University of Washington, Seattle, Washington 98195, USA}
\affiliation{Wayne State University, Detroit, Michigan 48201, USA}
\affiliation{Institute of Particle Physics, CCNU (HZNU), Wuhan 430079, China}
\affiliation{Yale University, New Haven, Connecticut 06520, USA}
\affiliation{University of Zagreb, Zagreb, HR-10002, Croatia}

\author{B.~I.~Abelev}\affiliation{University of Illinois at Chicago, Chicago, Illinois 60607, USA}
\author{M.~M.~Aggarwal}\affiliation{Panjab University, Chandigarh 160014, India}
\author{Z.~Ahammed}\affiliation{Variable Energy Cyclotron Centre, Kolkata 700064, India}
\author{B.~D.~Anderson}\affiliation{Kent State University, Kent, Ohio 44242, USA}
\author{D.~Arkhipkin}\affiliation{Particle Physics Laboratory (JINR), Dubna, Russia}
\author{G.~S.~Averichev}\affiliation{Laboratory for High Energy (JINR), Dubna, Russia}
\author{J.~Balewski}\affiliation{Massachusetts Institute of Technology, Cambridge, MA 02139-4307, USA}
\author{O.~Barannikova}\affiliation{University of Illinois at Chicago, Chicago, Illinois 60607, USA}
\author{L.~S.~Barnby}\affiliation{University of Birmingham, Birmingham, United Kingdom}
\author{J.~Baudot}\affiliation{Institut de Recherches Subatomiques, Strasbourg, France}
% ...

\collaboration{STAR Collaboration}\noaffiliation


\date{\today}
\begin{abstract}

We report the measurement of charged \Dstar mesons in inclusive jets produced in proton-proton collisions at a center of mass energy \srt = 200\,GeV with the STAR experiment at RHIC.
For \Dstar mesons with fractional momenta $0.2 < z < 0.5$ in inclusive jets with 11.5\,GeV mean transverse energy, the production rate is found to be $N(D^{*+}+D^{*-})/N(\mathrm{jet}) = 0.015 \pm 0.008 (\mathrm{stat}) \pm 0.007 (\mathrm{sys})$.
This rate is consistent with perturbative QCD evaluation of gluon splitting into a pair of charm quarks and subsequent hadronization.

\end{abstract}
\pacs{13.85.Ni, 13.87.Fh, 13.25.Ft}
\maketitle

Charm and bottom quarks can probe the partonic matter produced in heavy-ion collisions~\cite{STARwhitepaper} and the nucleon spin structure in polarized proton-proton collisions~\cite{RHICSpinCharm}. Their production mechanism is, therefore, of considerable interest at the Relativistic Heavy Ion Collider (RHIC).
Studies of the \Dstarpm-meson content in jets by the ALEPH, L3 and OPAL Collaborations~\cite{LEPdstar} show that the production from $Z^0$ decays in $e^++e^-$ collisions is dominated by \Dstar mesons that carry large fractions of the jet momenta, consistent with the jets being produced from primary $c$ (anti-)quarks.
The E531 and NOMAD Collaborations observed events with large $D^{*+}$ momentum fractions in neutrino charged-current interactions~\cite{CCdstar}.
In $\bar{p}+p$ collisions at \srt = 630\,GeV and 1.8\,TeV, the UA1 and CDF Collaborations have observed  \Dstarpm mesons in jets with transverse energies larger than 40\,GeV~\cite{UA1dstar,CDFdstar}.
Their fractional momenta are found smaller, consistent with a different production mechanism in which the \Dstar mesons originate from gluon splitting into $c\bar{c}$ pairs ($g\rightarrow c\bar{c}$ in the initial or final parton shower, with neither of the quarks from the $c\bar{c}$ pair participating in the hard QCD interaction)~\cite{gSplit}.
The multiplicity of such heavy quark pairs in gluon jets is calculable in perturbative QCD (pQCD), and the leading nonperturbative correction is believed to be small~\cite{pQCDsplit}.
At \srt = 200\,GeV, the RHIC energy, heavy quarks can still be produced via gluon splitting.
Perturbative QCD suggests that these contributions are small, and that the majority of the heavy quarks originate from gluon-gluon fusion~\cite{Nason:1987xz,Beenakker:1988bq}.
These expectations, however, have not until now been confronted with data at RHIC.

In this paper we present the first measurement of charged \Dstar mesons in inclusive jets produced in \pp collisions at a center of mass energy \srt = 200 GeV at RHIC.
The data were recorded in the year 2005 with the Solenoidal Tracker At RHIC (STAR)~\cite{starnim} and amount to an integrated luminosity of 2 pb$^{-1}$.
The main subsystems used in the measurement were the Time Projection Chamber (TPC) and the Barrel Electro-magnetic Calorimeter (BEMC), both located in a 0.5\,T solenoidal magnetic field.
The TPC provided tracking for charged particles with pseudo-rapidities $|\eta|\lesssim1.3$ for all azimuthal angles $\phi$.
The BEMC provided triggering and was used to measure photons and electrons.
In 2005 it covered $0<\eta<1$ in pseudo-rapidity and 2$\pi$ in azimuth.
Events used in this analysis were required to satisfy both a minimum bias trigger condition and a jet patch (JP) trigger condition. 
The minimum bias trigger was defined as a coincidence between Beam-Beam Counters (BBC) on either side of the interaction region, and the JP trigger, used also in Ref.~\cite{STARjet2005}, required the sum of transverse energies deposited in one of six $\Delta\eta\times\Delta\phi = 1\times 1$ patches of BEMC towers to be above a threshold of 6.5 GeV.

The charged \Dstar candidates were identified through the decay sequence $D^{*+}\rightarrow D^0\pi^+_s$,~$D^0\rightarrow K^-\pi^+$ and its charge conjugate. The \Dstar decay has a small $Q$-value. 
The $D^0$ thus carries most of the $D^{*}$ momentum and the pion from the $D^{*}$ decay, denoted by $\pi_s$, is soft. In the following, we will use $D^*$ to denote $D^{*+}+D^{*-}$, and $D^0$ to denote $D^0+\bar{D}^0$ unless specified otherwise.
The enhancement in the distribution of the invariant mass difference $\Delta M = M(K^{\mp}\pi^{\pm}\pi_s^{\pm}) - M(K^{\mp}\pi^{\pm})$ is used to determine the \Dstar yield~\cite{Nussinov:1975ay}. The candidate daughter kaons and pions were tracked with the TPC and, where possible, identified using the agreement of the measured and expected ionization energy loss ($dE/dx$) in the TPC to within two standard deviations.
The reconstructed tracks were required to have transverse momenta $p_{T}>0.2$ GeV/$c$ and pseudorapidities  $|\eta|<1$.
Only those events whose reconstructed primary interaction vertices were on the beam axis within 100 cm from the TPC center were retained.
A mass interval 1.82 $<M(K^{\mp}\pi^{\pm})<$ 1.90 GeV/$c^2$ was used to select $D^0$ candidates, consistent with the $D^0$ mass~\cite{pdg} and the experiment invariant mass resolution. About 90\% of $D^0$ signals are within this mass interval.
Combinatorial background was suppressed using the low $Q$-value of the $D^{*}$ decay by requiring the ratio, $r$, of the transverse momenta of the $D^0$ candidates and the soft pions to be $10 < r < 20$.
In addition, the decay angle of the kaon in the $K^{\mp}\pi^{\pm}$ rest frame, $\theta^*$, was restricted by requiring $\cos(\theta^*)<0.8$ to remove near-collinear combinatorial background from jet fragmentation.

\begin{figure}[ht]
\centerline{
\includegraphics[width=0.425\textwidth] {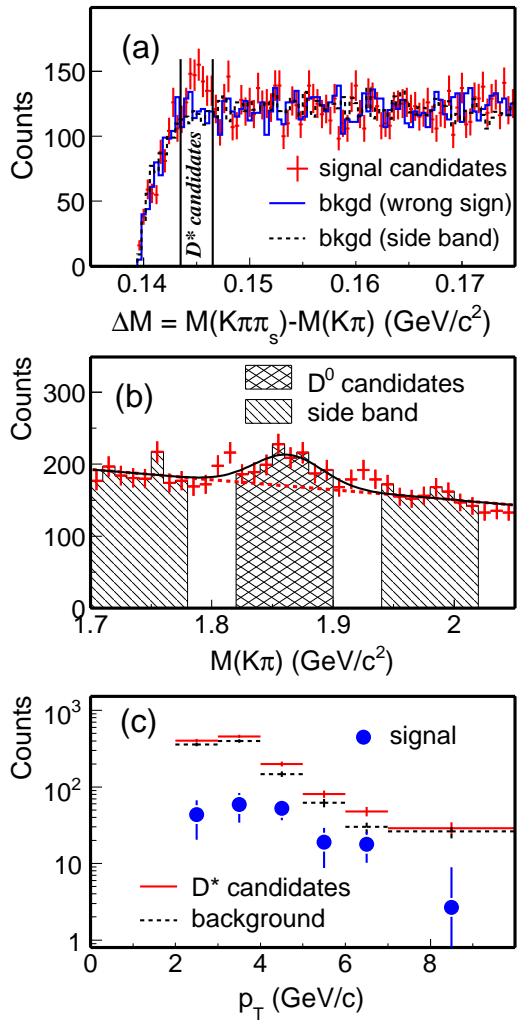}}
 \caption[]{(a) The observed distribution of the invariant mass difference $\Delta M = M(K^{\mp}\pi^{\pm}\pi_s^{\pm}) - M(K^{\mp}\pi^{\pm})$ in $p+p$ collisions at $\sqrt{s}$ = 200 GeV. The crosses show signal and background, and the histograms show two evaluations of the background, discussed in the text. (b) The invariant mass distribution of the $K^{\mp}\pi^{\pm}$ pairs for the events with an additional soft pion in the mass region 143.5 $< \Delta M <$ 146.5 MeV/$c^2$. The cross-hatched area depicts the $D^0$ mass interval used in selecting the \Dstar candidates and the hatched areas are used in constructing the background.  (c) The transverse momentum distribution of the \Dstar candidates after background subtraction. No corrections were applied for efficiency and acceptance.  The lower $p_T$ bound in this spectrum results from kinematic selection criteria applied to the \Dstar decay daughters.}
 \label{fig1mass}
\end{figure}

Figure ~\ref{fig1mass} (a) shows the spectrum of the invariant mass difference $\Delta M = M(K^{\mp}\pi^{\pm}\pi_s^{\pm}) - M(K^{\mp}\pi^{\pm})$.
The ``right sign'' combinations $K^{\mp}\pi^{\pm}\pi_s^{\pm}$ were used in obtaining the \Dstar candidates, while the doubly Cabbibo-suppressed ``wrong sign'' combinations $K^{\pm}\pi^{\mp}\pi_s^{\pm}$ were used as a measure of combinatorial background.
The ``wrong sign'' distribution in Fig.~\ref{fig1mass} (a) was superimposed directly on the ``right sign'' distribution, that is, without applying a relative normalization.
The hatched sidebands in the $K^{\mp}\pi^{\pm}$-mass spectrum of Fig.~\ref{fig1mass} (b) were used in an alternative measure of the \Dstar combinatorial background.
In the $K^{\mp}\pi^{\pm}$-spectrum, a subsample of all $K^{\mp}\pi^{\pm}$ candidates was chosen by requiring the event to contain an additional soft pion of the right charge sign resulting in 143.5 $<\Delta M<$ 146.5 MeV/$c^2$. This mass range contained about 95\% of the \Dstar signal.
The sample of all $K^{\mp}\pi^{\pm}$ candidates has considerably larger background and does not show a significant $D^0$ signal.
The crosshatched area underneath the $D^0$ peak indicates the mass interval used in the reconstruction of \Dstar candidates.
The sideband-based distribution of combinatorial background was normalized to the \Dstar signal candidate distribution in the mass region, 150 $<\Delta M<$ 175 MeV/$c^2$, well away from the \Dstar signal.
The combinatorial background distributions from the two methods are in good agreement.
The signal above background in Fig.~\ref{fig1mass} (a) has $\sim4\sigma$ significance and corresponds to  $180\,\pm\,45$ \Dstar counts.
The $D^{*+}$ and $D^{*-}$ yields of $96\,\pm\,32$ and $84\,\pm\,33$ counts are equal to within their statistical uncertainties, as expected at this level of precision~\cite{Braaten:2002yt}.
The sideband-based background distribution was used in extracting these yields since it results in better precision.
The difference with the wrong-sign results was used in assessing systematic uncertainties of the measurement. 
Raw $p_T$ distribution is shown in Fig.~\ref{fig1mass} (c).

Jets were reconstructed using a mid-point cone algorithm~\cite{midcone} which clusters reconstructed TPC tracks and  BEMC energy deposited within a cone in $\eta$ and $\phi$ of radius $r_\mathrm{cone} = 0.4$, as described in Refs.~\cite{STARjet,STARjet2005}. 
Events with reconstructed primary interaction vertex positions on the beam axis within 100 cm of the TPC center were kept for further analysis.
Jets were required to have \pt $>$ 8 GeV/$c$, $0<\eta<1$, and an electro-magnetic fraction of the jet transverse energy within 0.1 and 0.9 to reduce the effects of event pile-up and beam background~\cite{STARjet2005}.
A sample of $1.7\times10^6$ jets that pointed to a triggered jet patch was retained.

\begin{figure}[floatfix] \centerline{\includegraphics[width=0.45\textwidth]
{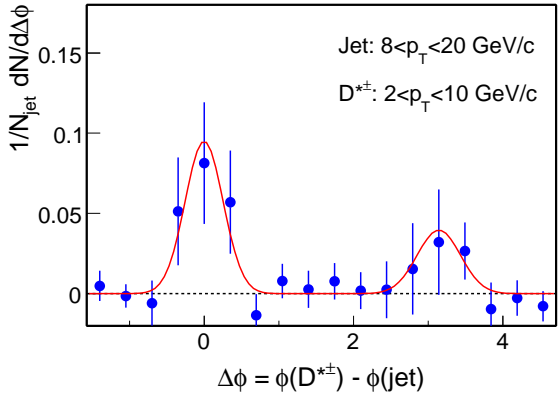}} \caption[]{The distribution of the \Dstar azimuthal angle with respect to the reconstructed jet axis from \pp JP triggered data. The distribution has been corrected for the \Dstar reconstruction efficiency. The curve is a two-Gaussian fit to the data points.
} \label{fig2phicorr}
\end{figure}

Figure~\ref{fig2phicorr} shows the distribution of the {\Dstar}-candidate azimuthal angle with respect to the reconstructed jet axis. 
Background was subtracted using the sideband method.
The distribution was corrected for the \Dstar reconstruction efficiency and acceptance obtained from \textsc{pythia}-based (v 6.205~\cite{pythia} `CDF TuneA' settings~\cite{CDF:TuneA}) Monte Carlo simulations passed through \textsc{geant}-based~\cite{geant:321} STAR detector response simulation.
The same simulation setup has been used in Refs.~\cite{STARjet,STARjet2005} and provides an adequate description of the inclusive jet data.
The determination of the \Dstar reconstruction efficiency took into account different configurations where some or all of the \Dstar decay daughters were part of the reconstructed jet and also different intervals for the jet and \Dstar momenta were taken into account.
The indicated uncertainties are the quadratic sum of the statistical uncertainties in the data and in the Monte-Carlo simulation. The solid line is a two-Gaussian fit to the data points.
A clear correlation is observed at the near side as expected.
The away side correlation is limited by statistics.
In the following we will focus on the near side correlation to investigate the production of charm in jets.

\begin{figure}[floatfix] \centerline{\includegraphics[width=0.45\textwidth]
{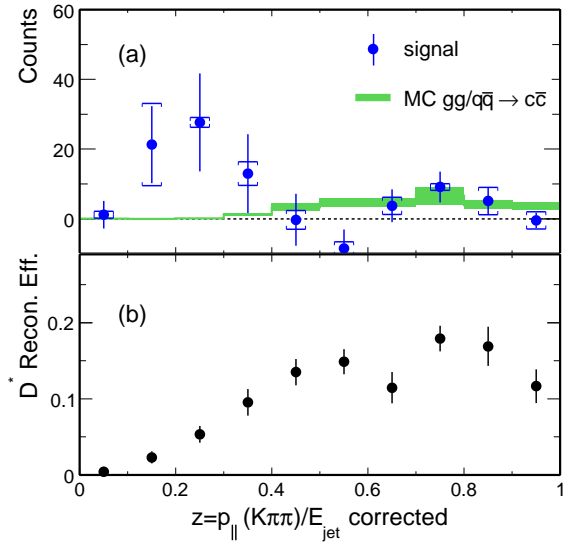}} \caption[]{ (a): The distribution of the \Dstar longitudinal momentum fraction $z$ in jets from JP triggered data. The size of the statistical uncertainties are indicated by the bars and the size of the background-subtraction systematic by the brackets. No corrections were applied for trigger effects and \Dstar reconstruction efficiency; however, the observed jet momenta and hence $z$ were corrected for the detector response. The data at large $z$ are compared with a Monte Carlo simulation of charm creation through $gg/q\bar{q}\rightarrow c\bar{c}$. (b): The average \Dstar reconstruction efficiency versus $z$. 
} \label{fig3zcorr}
\end{figure}

For near side \Dstar candidates, the fragmentation variable $z\equiv p_{\parallel}(K^{\mp}\pi^{\pm}\pi_s^{\pm})/E_{\mathrm{jet}}$ was computed, where $p_{\parallel}(K^{\mp}\pi^{\pm}\pi_s^{\pm})$ is the $K^{\mp}\pi^{\pm}\pi_s^{\pm}$ momentum projection on the jet axis, and $E_{\mathrm{jet}}$ is the jet total energy.
The reconstructed jet transverse energy is on average about 20\% larger than the generated jet transverse energy, mostly because of the sharply decreasing jet yield with increasing transverse energy and the jet transverse energy resolution.
The resolution has been studied by Monte-Carlo simulation and by using transverse momentum balance in a sample of dijet events~\cite{STARjet,STARjet2005}.
The reconstructed jet transverse energy has been corrected and residual effects are accounted for in the systematic uncertainties.
The distribution of $z$, after this correction, is shown in Fig.~\ref{fig3zcorr}(a). The signal in this spectrum corresponds to 72 $\pm$ 25 counts.
The uncertainties represented by the bars are statistical and the brackets indicate the contribution caused by combinatorial background subtraction.
No corrections were made here for trigger effects and reconstruction efficiency. The average \Dstar $p_T$ is $\sim$3 GeV/$c$ for $0.2<z<0.5$, and $\sim$6 GeV/$c$ for $z>0.5$.
The average \Dstar reconstruction efficiency from simulation, shown in Fig.~\ref{fig3zcorr} (b), is found to increase with increasing $z$.
The trigger efficiency largely cancels in the measurement of the production ratio $N(D^*)/N(\mathrm{jet})$ of interest here. However, the JP trigger condition preferentially selects jets with large electromagnetic energy. It thus disfavors jets containing the hadronic decay products of the \Dstar mesons, in particular for high $z$. The effects of this trigger bias were studied by comparing the simulated jet yields with and without the JP trigger condition. 
Their ratio is found constant below $z\!\sim\!0.5$ and decreases rapidly for larger $z$, as expected.
The green band in Fig.~\ref{fig3zcorr} (a) was obtained by simulating only the direct charm flavor creation processes, $gg\rightarrow c\bar{c}$ and $q\bar{q}\rightarrow c\bar{c}$, in \textsc{pythia} and passing the results through the STAR detector response simulation.
The simulated data were analyzed in the same way as the real data and were normalized using the measured total charm production cross section~\cite{Adams:2004fc}. Only a small faction of the generated events containing \Dstar mesons with $z>$ 0.5 satisfies the JP trigger condition. To within the large uncertainties good agreement is found with the \Dstar data at high $z$, where the production of charmed hadrons is expected to be dominated by charm quark fragmentation.
The excess observed in the data at smaller $z$ can be ascribed to production processes that are not included in the simulation, such as gluon splitting.

The ratio $N(D^*)/N(\mathrm{jet})$ was determined for the region $0.2 < z < 0.5$. For $z<0.2$, the \Dstar reconstruction efficiency is low, and for $z>0.5$, the JP trigger is strongly biased against jets with \Dstar mesons that decay into charged hadrons.
After correcting for the \Dstar reconstruction efficiency, shown in Fig.~\ref{fig3zcorr} (b), and the decay branching ratio of $(67.7 \pm 0.5)$\% for $D^{*+}\rightarrow D^0\pi^+_s$ and of $(3.89 \pm 0.05)$\% for $D^0\rightarrow K^-\pi^+$~\cite{pdg}, we obtain \Nratio = $0.015\pm0.008(\mathrm{stat})\pm0.007(\mathrm{sys})$ for $0.2<z<0.5$ and a mean jet transverse energy of \met = 11.5 GeV.
The estimated statistical uncertainty includes the statistical uncertainty in the simulations. 
The main contributions to the systematic uncertainty are estimated to originate from the jet definition and selection ($\sim$35\%), from trigger bias ($\sim$18\%), from \Dstar combinatorial background ($\sim$10\%), and from the \Dstar reconstruction efficiency ($\sim$10\%).
The uncertainties associated with the jet definition and selection were estimated by varying the accepted  primary vertex range, the jet $\eta$ range, and the criteria used to reduce the effects of event pile-up and beam background.
The effects from trigger bias were assessed by Monte-Carlo simulation. The size of the background uncertainty was estimated by comparing the results obtained with the different background subtraction methods.
The uncertainty in the \Dstar reconstruction efficiency was estimated by varying the daughter particle track quality criteria.
The contributions were combined in quadrature to obtain the total systematic uncertainty estimate.

\begin{figure}[h]
\centerline{\includegraphics[width=0.45\textwidth]{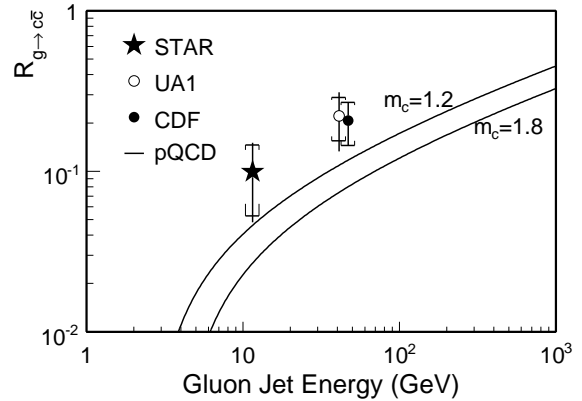}}
\caption[]{Gluon splitting rate to charm pairs as a function of the gluon jet energy. Measurements from STAR, UA1~\cite{UA1dstar} and CDF~\cite{CDFdstar} collaborations are compared with pQCD calculations~\cite{pQCDsplit} using the indicated values of the charm quark mass (in GeV/$c^2$), $\Lambda_\mathrm{QCD} = 300\,\mathrm{MeV}$, and a Peterson fragmentation function with $\epsilon_c = 0.06$.} \label{fig4e}
\end{figure}

To estimate the rate of gluon splitting into charm pairs, $R_{g\rightarrow c\bar{c}}$, from the ratio $N(D^*)/N(\mathrm{jet})$ one needs to correct for the unmeasured $z$ region, the fraction of charm quarks that fragment into $D^*$, and the fraction of gluon jets in the sample. The fraction of gluon jets in the data was estimated to be 60\% from \textsc{pythia} simulations and from next-to-leading order pQCD evaluation~\cite{vogelsang}.
A 10\% uncertainty is included as a systematic contribution in $R_{g\rightarrow c\bar{c}}$.
The $c\rightarrow D^{*+}$ and $\bar{c}\rightarrow D^{*-}$ fraction is taken to be $(22.4\pm2.8)\%$~\cite{pdg}.
This is smaller than the value of 3/8 estimated in the earlier publications by UA1~\cite{UA1dstar} and CDF~\cite{CDFdstar}.
By using the leading-order pQCD evaluation of gluon splitting~\cite{pQCDsplit}, we estimate that the measured ratio $N(D^*)/N(\mathrm{jet})$ for $0.2<z<0.5$ captures $(53\pm5)\%$ of $R_{g\rightarrow c\bar{c}}$ at \met = 11.5 GeV and the dominant part of the remainder resides at smaller $z$.
This percentage was then used to extrapolate over the unmeasured $z$ region.
Our result for $R_{g\rightarrow c\bar{c}}$ is shown in Fig.~\ref{fig4e}, together with the UA1 and CDF measurements~\cite{UA1dstar,CDFdstar}.
The results are compared to a theoretical evaluation in leading-order pQCD~\cite{pQCDsplit}.
The expectation is consistent with the data to within the combined experiment statistical and systematic uncertainties.
Although the agreement is not strong, the conclusion that $R_{g\rightarrow c\bar{c}}$ is small for energies accessible at RHIC is clearly supported.
The use of the PDG estimate for the fraction $c\rightarrow D^{*+}$~\cite{pdg} for the UA1 and CDF data does not change this conclusion.

The pQCD expectation for $R_{g\rightarrow c\bar{c}}$~\cite{pQCDsplit} can be combined with the STAR measured mid-rapidity jet differential cross section~\cite{STARjet}, covering jet $p_T$ in the range of 5 to 50 GeV/$c$, and the gluon jet fraction~\cite{vogelsang} to estimate the gluon splitting contribution to the charm production cross section. The gluon splitting contribution can in this way be determined for charm $p_T$ in the range of 2 to 10 GeV/$c$.
The result is smaller by an order of magnitude than the pQCD expectation~\cite{FONLL} for the total charm differential production cross section at RHIC.
It is thus small also compared to the measured total charm cross section~\cite{Adams:2004fc}, which exceeds the pQCD expectation.



In summary, we report the first measurement on the charm content in jets from \pp collisions at \srt = 200 GeV.
The ratio $N(D^{*+}+D^{*-})/N(\mathrm{jet})$ is measured to be $0.015 \pm 0.008 (\mathrm{stat}) \pm 0.007 (\mathrm{sys})$ for \Dstar mesons with fractional momenta $0.2 < z < 0.5$ in jets with a mean transverse energy of 11.5\,GeV.
This is consistent with perturbative QCD evaluation of gluon splitting into a pair of charm quarks and subsequent hadronization into \Dstar mesons.
The associated cross section is smaller than the total charm production cross section at RHIC.
We thus infer that the charm content in jets at RHIC energies has a small contribution from gluon splitting and is dominated by jets initiated by charm quarks.

We thank the RHIC Operations Group and RCF at BNL, and the NERSC Center 
at LBNL and the resources provided by the Open Science Grid consortium 
for their support. This work was supported in part by the Offices of NP 
and HEP within the U.S. DOE Office of Science, the U.S. NSF, the Sloan 
Foundation, the DFG cluster of excellence `Origin and Structure of the 
Universe', CNRS/IN2P3, RA, RPL, and EMN of France, STFC and EPSRC 
of the United Kingdom, FAPESP of Brazil, the Russian Ministry of Sci. 
and Tech., the NNSFC, CAS, MoST, and MoE of China, IRP and GA 
of the Czech Republic, FOM of the Netherlands, DAE, DST, and CSIR 
of the Government of India,  the Polish State Committee 
for Scientific Research,  and the Korea Research Foundation.


\end{document}